\documentclass[twocolumn,reprint,showpacs,preprintnumbers,aps,prl,letterpaper,superscriptaddress,nofootinbib,tightenlines,floats,floatfix]{revtex4-1}
\usepackage{graphicx}
\usepackage{dcolumn}
\usepackage{bm}
\usepackage{latexsym}
\usepackage{amsmath,amssymb}
\usepackage[draft=false]{hyperref}
\usepackage[latin1]{inputenc}
\usepackage{latexsym}
\usepackage{amsmath}
\usepackage{ifpdf}
\usepackage{color}
\usepackage{epstopdf}
\usepackage{subcaption}
\usepackage[skip=2pt,font=scriptsize]{caption}

\begin{document}
\title{Exact Rotating Magnetic Traversable Wormholes satisfying the Energy Conditions}

\author{Galaxia Miranda}
\email{mmiranda@fis.cinvestav.mx}
\affiliation{Departamento de F\'isica, Centro de Investigaci\'on y de
  Estudios Avanzados del IPN, A.P. 14-740, 07000 M\'exico D.F., M\'exico.}
\author{Tonatiuh Matos}
\email{tmatos@fis.cinvestav.mx}
 \altaffiliation{Part of the Instituto Avanzado de Cosmolog\'ia (IAC)
  collaboration http://www.iac.edu.mx/}
\affiliation{Departamento de F\'isica, Centro de Investigaci\'on y de
  Estudios Avanzados del IPN, A.P. 14-740, 07000 M\'exico D.F.,
  M\'exico.}

\begin{abstract}
In this work we wonder if there is a way to generate a wormhole (WH) in nature using ``normal" matter. In order to give a first answer to this question, we study a massless scalar field coupled to an electromagnetic one (dilatonic field) with an arbitrary coupling constant, as source of gravitation. We obtain an exact solution of the Einstein equations using this source that represents a magnetized rotating WH. This space-time has a naked ring singularity, probably untouchable as in \cite{Matos:2012gj}, but otherwise regular. The WH throat lies on the disc bounded by the ring singularity, which keeps the throat open without requiring exotic matter, that means, satisfying all the energy conditions. After analyzing the geodesic motion and the tidal forces we find that a test particle can go through the WH without troubles.
\end{abstract}

\date{Received: date / Accepted: date}

\maketitle

\date{\today}


\maketitle

\section{Introduction}

In 1916, Flamm suggested that our Universe might not be simply connected \cite{Flamm}, opening the possibility to the existence of tunnels connecting different regions of the universe or even completely different universes. In 1935 Einstein and Rosen rediscover this solution trying to give a field representation of particles \cite{ER}, idea further by Ellis \cite{Ellis}, who models particles as bridges between two regions of the space-time. Many years latter Morris and Thorne consider such solutions as means of interstellar travel \cite{MT}. Nevertheless, these wormholes (WH) need to violate the energy conditions. Today this type of matter is called exotic (see \cite{Viser1} for a detailed review on this subject).

In \cite{Hochberg97,Hochberg98} it was showed that the violation of the null energy condition was a generic feature for regular traversable WHs, assuming that the throat is a compact $2D$ surface with minimum area. However, if the throat is no longer a compact object (i.e. a star-like structure) one might deal with cylindrical WHs, which from afar appear as cosmic strings, without requiring the presence of exotic matter \cite{Bronn14,Bronn13,Clem95}. 

On the other side, numerical simulations seem to show that static WHs are unstable \cite{paco1}. To flip it this problem, it was conjectured in \cite{dario} that the rotation of the WH could stabilize a ghost star. The idea is that a rotating WH would have more possibilities to be stable than the general static spherically symmetric WHs. Some rotating solutions were studied in the past, as an approximation \cite{rot,rotk} or as an exact solution of the Einstein equations \cite{dario, mio, Matos:2012gj}. However, all of them violet some energy condition. 

Nevertheless, we wonder if it is possible to generate WHs where the source is some kind of matter that can be found in nature.
In this work we look for WHs made of normal matter that are traversable, that is, WH made of matter that satisfy the energy conditions and where a test particle can go from one side of the throat to the other in a finite time without facing large tidal forces.  We also look for rotating WHs, following the conjecture that the rotation can stabilize the WH.  

In order to give an answer to this question, at least partially, we look for scalar fields that could be formed by particles coupled with the electromagnetic field. Of course this kind of particles exist and are common in nature, but all of them are massive. The problem is that so far it is not possible to get exact solutions of the Einstein equation with massive scalar fields.
However, if the scalar field is massless it is possible to use standard techniques to find exact space-times from this sources. We lost some important feature of the matter but we gain precision to see the form of the space-time here, which is the most important point for the analysis of this kind of space-times. Thus, we start from the Lagrangian
\begin{equation}
L=-R+2\varepsilon(\nabla\Phi)^2+e^{-2\alpha\Phi}F^2+V(\Phi)
\label{eq:lagrangian}
\end{equation}
where $R$ is the Ricci scalar, $F_{\mu\nu}$ is the Faraday tensor, $\Phi$ is the field of a spin cero (composed) particle and $V(\Phi)$ the scalar field potential. We separate the dilatonic from the ghost field using $\varepsilon=+1$ for a dilatonic and $\varepsilon=-1$ for the ghost field. As a first approximation we set the mass of the scalar field equal to cero $V=0$, as we will see, we can solve the Einstein equation exactly, giving us a space-time that we can study with all the precision.  
The Einstein-Maxwell-Dilaton field equations from the Lagrangian (\ref{eq:lagrangian}) are
\begin{equation}
R_{\mu \nu}=2 \, \varepsilon \Phi_{,\mu} \Phi_{,\nu} + 2 \, e^{-2 \alpha \Phi} \left( F_{\mu \rho} F_{\nu}^{\, \, \rho} - \frac{1}{4} g_{\mu \nu} F_{\delta \gamma} F^{\delta \gamma} \right) 
\label{eq:EFE}
\end{equation} 
%

In this work we report an asymptotically flat rotating magnetized solution of the Einstein equation obtained using the ansatz proposed in \cite{phantom}. 
 The scalar field $\Phi$ is given by
\begin{equation}
 \Phi=\frac{a \, y}{\alpha L^2 \,(x^2+y^2)}\ .
\label{eq:Phi}
\end{equation}
and the vector potential
\begin{eqnarray}
A_{\mu} =-\frac{e^{\alpha \Phi}}{2} \, \left[ 1- e^{-\alpha \Phi} ,0,0,\frac{a\, x \,(1-y^2)}{L(x^2+y^2)} \right] ,
\label{eq:Amu}
\end{eqnarray}
in oblate spheroidal coordinates with $x \in \mathbb{R}$ and $|y|\leq 1$ 
\begin{eqnarray}
ds^2&=& L^2\,\left[(x^2+y^2)e^{K}\left( \frac{dx^2}{x^2+1}+\frac{dy^2}{1-y^2} \right)\right. \nonumber\\
&+&\left.(x^2+1)(1-y^2)d\varphi^2\right] \notag  \\ 
&-&\left(dt+\frac{a}{L}\frac{x \, (1-y^2)}{x^2+y^2}\,d\varphi\right)^2 ,
\label{eq:solBL}
\end{eqnarray}  
This space-time has a scalar charge parameter $a$. 
The scalar field (\ref{eq:Phi}) and the electromagnetic four potential (\ref{eq:Amu}) are solutions of the Einstein equations (\ref{eq:EFE}) provided that 
\begin{equation}
K=\frac{k}{L^4}\,\frac{(1-y^2)(8x^2\,y^2(x^2+1)-(1-y^2)(x^2+y^2)^2)}{(x^2+y^2)^4}
\end{equation}
with the relation for the free constants $\alpha$, $a$ and $k$ given by 
\begin{equation}
\alpha^2 (a^2 - 8k)-4\varepsilon\,a^2=0.
\end{equation}
Here $L > 0$ is a parameter with units of distance, $a$ is a parameter with units of angular momentum and $\alpha$ the coupling constant.
This space-time represents a magnetized rotating WH, without gravitational potential. The space-time is curved by the presence of the magnetic charge and the scalar field, while the magnetic field represents a magnetic dipole.

Interesting special cases are the dilatonic field with $\alpha^2=1$ (which represents a low-energy string theory), where we get $k=-\frac{3a^2}{8}$, while for the ghost field with the same coupling constant we obtain $k=\frac{5a^2}{8}$. If $\alpha^2=3$, for the dilatonic field (in this case the action reduces to a $5D$ Kaluza-Klein theory)  we obtain $k=-\frac{a^2}{24}$ and for the ghost field $k=\frac{7\,a^2}{24}$. If $k=0$ then $\alpha^2=4$, here only the dilatonic field with $\varepsilon=1$ is possible.

All the invariant quantities of metric (\ref{eq:solBL}), like the Ricci scalar, the quadratic Riemann tensor, etc. are of the form
\begin{equation}
\text{Invariants}=\frac{F(x,y)}{(x^2+y^2)^\beta} e^{-K}
\label{eq:invariantes}
\end{equation}
where $\beta$ is a positive integer and $F(x,y)$ is a polynomial of degree less than the degree of $(x^2+y^2)^\beta$ and of less order than $e^{-K}$. From (\ref{eq:invariantes}) we see that the space-time (\ref{eq:solBL}) has an anisotropic naked ring singularity of radius $L$ at $x=y=0$. 

On the other hand, (\ref{eq:solBL}) is asymptotically flat since 
\begin{equation}
\lim_{x\rightarrow \infty} e^{K}\rightarrow 1 \hspace{0.5cm} \text{and} \hspace{0.5cm} \lim_{x\rightarrow \infty} \frac{a}{L}\frac{x \, (1-y^2)}{x^2+y^2} \rightarrow 0 \nonumber
\end{equation}

The throat of the WH lie on the disc $x=0$ bounded by the singularity $y=0$. The throat connects two $3D$-spaces, one with $x>0$ and another with $x<0$, called the ring WH  \cite{Bronn10}. Since there are no discontinuities in the extrinsic curvature on the disc, it is possible to cross the surface $x=0$, i.e., to travel through the WH, since crossing this bounded surface represents leaving one world and entering another. 

%
It has been shown that for WHs whose throat is a regular compact $2D$ surface with a finite minimum area, the violation of the null energy condition (NEC) near or at the throat is required \cite{Hochberg98,Hochberg97}. 
In order to analyse the energy conditions, we choose an orthonormal basis \cite{kuf05} 
\begin{eqnarray}
e_{\hat{t}}&=&e_{t},\hspace{0.6cm} 
e_{\hat{x}}=\sqrt{\frac{x^2+1 }{L^2(x^2+y^2)e^{K}}}e_{x}, \\
e_{\hat{y}}&=&\frac{ e_{y} }{\sqrt{L^2 (x^2+ y^2)e^{K}}},\nonumber \hspace{0.2cm}
e_{\hat{\phi}}=\frac{e_{\phi}-\omega e_{t}}{\sqrt{m^2(x^2+1)(1-y^2)}}, 
\label{tetrad}
\end{eqnarray}
where $\omega=(a x(1-y^2))/(L(x^2+y^2))$, and $e_{\alpha}=\partial /\partial x^{\alpha}$ is the canonical vectors basis.
In the simplest scenario where $e^{K}=1$ with $\alpha=2$, we use an outgoing null vector in the $x$ direction, $\mu=e_{\hat{t}}\pm e_{\hat{x}}$, thus $T_{\hat{\alpha}\hat{\beta}}\mu^{\hat{\alpha}}\mu^{\hat{\beta}}=T_{\hat{t}\hat{t}}+T_{\hat{x}\hat{x}}=\frac{1}{2\,}(R_{\hat{t}\hat{t}}+R_{\hat{x}\hat{r}})$. Then

\begin{equation}
\rho-\tau = T_{\hat{t}\hat{t}}+T_{\hat{x}\hat{x}} =\frac{ a^2}{2 m^6}\frac{x^4+x^4y^2+2x^2y^4+y^4(1-y^2)}{(x^2+y^2)^5} > 0,
\label{NEC}
\end{equation}
where  $\rho=T_{\hat{t}\hat{t}}$ is the total energy density of mass-energy and $-\tau=T_{\hat{x}\hat{x}}$ is the tension per unit area measured by the static observers in the $x-$direction.
Of course, for $\varepsilon=1$ the NEC is satisfied, since $\rho > \tau$ everywhere. 

On the other hand the energy density is given by

\begin{equation}
\rho=\frac{a^2}{4 L^6}\frac{x^2(1+3 y^2) + y^2 (1-y^2)}{(x^2+y^2)^5} > 0,
\label{WEC}
\end{equation} 

Thus, for $\alpha=2$ the NEC and WEC are satisfied, there is no need for exotic matter to keep the throat open, the ring singularity is the responsible for the existence of the WH \cite{Bronn14,Clem95}.  The presence of exotic matter that violates NEC have been changed for the ring singularity.

In the general case, for the dilatonic scalar field with $\varepsilon=1$ and $\alpha$ arbitrary, taking the same outgoing null vector as previously, we obtain 
\begin{eqnarray}
\rho-\tau&=&q^2\frac{e^{K}}{2 L^6 (x^2+y^2)^5}(x^4+x^4y^2+2x^2y^4+y^4(1-y^2)\nonumber\\
&-&2\small{\frac{\alpha^2-4}{\alpha^2}}x^2y^2(x^2+1)) > 0,
\end{eqnarray} 
Of course, NEC is satisfied if a traveller moves along the $x$ direction in an outgoing null vector.

For the ghost field with $\varepsilon=-1$, 
there are always regions where NEC is violated.


Due to the presence of the ring singularity ($x=y=0$) it is possible that a traveller crossing the throat experiences strong gravitational forces. To see if the throat is traversable we analyze the tidal forces following \cite{MT,rot,kuf05}. We take the reference frame of a traveller moving in the $x$ direction
\begin{eqnarray}
e_{\hat{0}}&=&\gamma e_{\hat{t}}\mp \gamma(v/c)e_{\hat{x}}, \hspace{0.3cm}
e_{\hat{1}}=\mp\gamma e_{\hat{x}}+\gamma(v/c)e_{\hat{t}}, \nonumber\\
e_{\hat{2}}&=&e_{\hat{y}}, \hspace{0.3cm} 
e_{\hat{3}}=e_{\hat{\phi}}.
\end{eqnarray}
being $\gamma=[1-(v/c)^2]^{-\frac{1}{2}}$. 
In the $x$ direction the tidal constraint is given by
\begin{eqnarray}
|R_{\hat{1}\hat{0}\hat{1}\hat{0}}|\leq g_{\oplus}/c^{2}\times 2m\approx 1/(10^{5}km)^{2},
\end{eqnarray}
where the height of our traveller is $2\,m$. We have $|R_{\hat{1}\hat{0}\hat{1}\hat{0}}|=|R_{\hat{x}\hat{t}\hat{x}\hat{t}}|$. 
%
%
while the remaining constraints are reduced to the study of $|R_{\hat{2}\hat{0}\hat{2}\hat{0}}|\leq (10^{5}km)^{-2}$, and $|R_{\hat{3}\hat{0}\hat{3}\hat{0}}|\leq (10^{5}km)^{-2}$. 
Since our metric is axially symmetric, we have that
\begin{eqnarray}
|R_{\hat{2}\hat{0}\hat{2}\hat{0}}|=\gamma^{2}|R_{\hat{y}\hat{t}\hat{y}\hat{t}}|
+\gamma^{2}(v^{2}/c)|R_{\hat{y}\hat{x}\hat{y}\hat{x}}|.
\end{eqnarray}

Assuming the traveller is at rest at the throat \cite{MT}, this implies $v\rightarrow 0$ and $\gamma\rightarrow 1$. Then $|R_{\hat{2}\hat{0}\hat{2}\hat{0}}|=|R_{\hat{y}\hat{t}\hat{y}\hat{t}}|$. 
%
%
%
Finally our last constraint can be written as $|R_{\hat{3}\hat{0}\hat{3}\hat{0}}|=|R_{\hat{\phi}\hat{t}\hat{\phi}\hat{t}}|$.
%


The constraints, for both dilatonic and ghost scalar field, are given by

\begin{eqnarray}
|R_{\hat{x}\hat{t}\hat{x}\hat{t}}| &=&\frac{e^{-K}}{4} \frac{a^2 (1-y^2) (x^2-y^2)^2}{L^{6}(x^2+y^2)^5},
\label{TC1}
\end{eqnarray}

\begin{eqnarray}
|R_{\hat{y}\hat{t}\hat{y}\hat{t}}| &=& e^{-K} \frac{a^2 x^2 y^2(x^2+1)}{L^{6}(x^2+y^2)^5},
\label{TC2}
\end{eqnarray}

\begin{eqnarray}
|R_{\hat{\phi}\hat{t}\hat{\phi}\hat{t}}|&=& \frac{e^{-K}}{4} \frac{a^2(x^2+3x^2 y^2+y^2(1-y^2))}{L^6(x^2+y^2)^4}.
\label{TC3}
\end{eqnarray}

Forcing traveller to approach the throat with $y=1$, the tidal force in the $x$ direction is zero everywhere, while the remaining tidal forces go to zero as the traveller approaches the throat at $x=0$. It is possible to traverse the throat without feeling the presence of the ring singularity traveling on the plane $y=1$.




In what follows we study the geodesics. We first are interested in radial geodesics to see whether an observer can penetrate the WH or not. Of course the ring singularity apparently does not allow any observer to penetrate the WH, at least going by the equator.

For doing so, let $\lambda$ be an affine parameter and  $u^\mu=(\dot t,\dot x,\dot y,\dot \phi)$, with $ {\dot t}=\frac{d t}{d\lambda}$, etc., the vector velocity of an observer, such that the equation $u^\mu u_\mu=-\kappa$ holds, with $\kappa =0 $ for lightlike geodesics and $\kappa =-1$ for timelike geodesics. It follows
\begin{eqnarray}
&&\kappa=-\left(\dot t+\frac{a\, x(1-y^2)}{L( x^2+y^2)}\,\dot\phi\right)^2 \label{eq:LagranBL}\\
 &+&L^2\left[ (x^2+ y^2)e^{K}\left(\frac{\dot x^2 }{x^2+1}+\frac{\dot y^2}{(1-y^2)}\right)\right.\nonumber\\
 &+&\left.( x^2+1)(1-y^2)\dot\phi^2\right]. \nonumber
\end{eqnarray}
We first analyze the geodesics on the plane $y=0$. (\ref{eq:LagranBL}) reduces to
\begin{equation}
L^2 x^4 e^{K_0}\dot x^2= L^2 x^2(x^2+1)(\mathcal{E}^2 +\kappa)-(\mathcal{L} L x+a\,\mathcal{E})^2=\hat X(x)  \label{eq:LagranBL3}
\end{equation}
with $K_{0}=-k/(m^4 x^4)$ and constants of motion $\mathcal{E}=(\dot{t}+\omega \dot{\phi})$ and $\mathcal{L}+\omega\mathcal{E}=L^2(x^2+1)(1-y^2)\dot\phi$ . 
$\hat X(x) \geq 0$ dominates the geodesics on $y=0$ \cite{GeomKerr}. 
Since the last equation \eqref{eq:LagranBL3} always admits at least a positive real root ($x_+>0$) two types of motion are possible \cite{Zakharov}. 
\begin{enumerate}
 \item  If the right hand side polynomial has roots such that there is a maximum one with $x_{max}>0$ and $(\partial \hat X/ \partial x)(x_{max})>0$, the particle departs to infinity after approaching $x_{max}$. \\
 \item  If $\hat X(x)$ has a root $x_{max}$ and $\hat X(x_{max})=(\partial \hat X/ \partial x)(x_{max})=0$ then the particle takes an infinite proper time to approach $x_{max}$.
\end{enumerate}
Since $\hat X(x=0)=-a^2 \mathcal{E}^2<0$, it is not possible for any geodesic to go through the WH throat or to reach the singular ring from the plane. On the other hand, if a test particle starts its motion on one side of the throat with $y=0$, this particle is going to remain on the same side of the throat all the time. Thus, the disc cannot be crossed on the plane $y=0$ due to the presence of the ring singularity.


On the other side, observe that on the plane $x=0$ with $\dot{\phi}=0$, the motion is governed by
\begin{equation}
\dot y^2=\frac{(\kappa+\mathcal{E}^2)(1-y^2)}{L^2 y^2}e^{K^0} \label{geoY}
\end{equation}
where now $K^0=\frac{-k(1-y^2)^2}{L^4 y^4}$. From (\ref{geoY}) and $\dot{t}=\mathcal{E}$, we obtain 
\begin{equation}
 dy=\frac{\sqrt{\kappa+\mathcal{E}^2}}{\mathcal{E} L}\frac{\sqrt{1-y^2}}{ y}e^{-K^0/2} dt \label{tau}
\end{equation}
as $y\rightarrow 0$ we get
\begin{eqnarray}
\frac{dy}{dt} \rightarrow 0 \hspace{0.2cm} \text{if} \hspace{0.2cm} k<0, \hspace{0.3cm}
\frac{dy}{dt} \rightarrow \infty \hspace{0.2cm} \text{if} \hspace{0.2cm} k\geq0
\end{eqnarray}
That is, for the dilatonic field with $\alpha^2 \geq 4$ and the ghost field with $\alpha>0$ the velocity increases infinitely as we approach the singularity. While, for a dilatonic field with $\alpha^2<4$, the test particles tend to be at rest as $y \rightarrow 0$.  For $k\geq 0$, a light ray on $x=0$ will rotate rapidly, increasing its velocity until it reaches $y=0$. If $k<0$, i.e. $\alpha=\sqrt{3}$, the same light ray will decrease its velocity as $y\rightarrow 0$ and eventually stop.  

\begin{figure}[htp]
\includegraphics[scale=0.7]{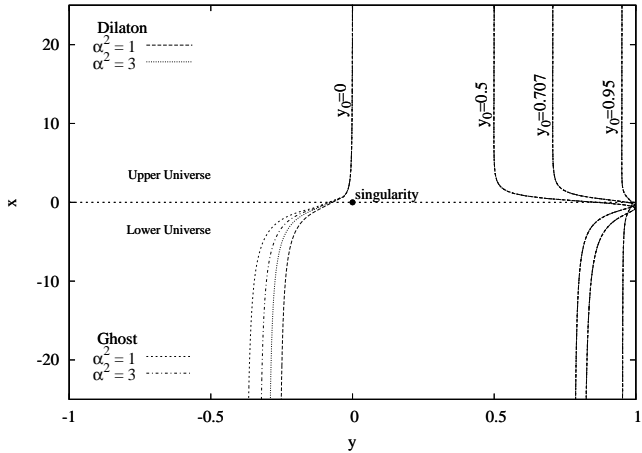}
\caption{Geodesics crossing the WH through the throat at $x=0$. The throat size is $L=10$ and the traveller initially at $(y_0,x_0=25)$. With $\mathcal{E}=10$, $\mathcal{L}=5$ and $a=0.1$, with two different values of $\alpha$.}
\end{figure}

Now we analyze a slowly rotating WH. At first order, in a slowly-rotating WH with $a^2/L^4 \rightarrow 0$,  the tidal forces (\ref{TC1})-(\ref{TC3}), the energy conditions (\ref{NEC})-(\ref{WEC}) and the Kretchmann tensor, approach to zero except at the ring singularity. In this limit it is possible to find a fourth conserved quantity $\mathcal{K}$. Using the Hamilton-Jacobi we get that the geodesic equations for $x$ and $y$ are given by
\begin{eqnarray}
L^2 \Delta^2 \dot x^2 = X(x), \hspace{0.5cm}
\Delta^2 \dot y^2 = Y(y), \nonumber 
\end{eqnarray}
where $\Delta =L^2(x^2+y^2)$, \hspace{0.1cm}  $\Delta_x =L^2(x^2+1)$, and
\begin{eqnarray}
X(x)&&=\Delta_x(L^2(\kappa+\mathcal{E}^2)x^2-\mathcal{K})-2aL\mathcal{EL}x+L^2\mathcal{L}^2,\label{X}\\
Y(y)&&=(1-y^2)(\mathcal{K}+(\kappa+\mathcal{E}^2)L^2y^2)-\mathcal{L}^2.\label{Y} 
\end{eqnarray}

In order to go through the WH, we need that $X(x)>0$ at $x=0$. From (\ref{X}), it follows that 
$
\mathcal{K} < \mathcal{L}^2, \label{cond1}
$
and forcing the geodesics to stay on the plane $y=y_0$, where $y_0$ is a non-zero constant, (\ref{Y}) must vanish. Consequently, from this two conditions   
\begin{equation}
\mathcal{K}<\mathcal{L}^2<(\kappa+\mathcal{E}^2)L^2(1-y_0^2)<(\kappa+\mathcal{E}^2)L^2\label{cond2}.
\end{equation}

If $y_0=1$, then $\mathcal{L}=0$ and from (\ref{cond2}) we get that $\mathcal{K}< 0$.
In general, for geodesics with $y\neq y_0$ (\ref{cond2}) holds.


Conclusions. Metric (\ref{eq:solBL}) contains a ring singularity at $x=y=0$ of radius $L$. The disc $x=0$ can be identified with a throat surrounded by the ring singularity $y=0$. Due to the presence of this singularity we showed that NEC and WEC are satisfied for $\varepsilon=1$. 

Despite this singularity, in the slow-rotating limit, geodesics on the plane $y=1$ can cross (back and forth) the disc and travel through the throat to another asymptotically flat space-time without facing extreme tidal forces. It is also important to note that geodesics avoid the ring singularity approaching the throat from $y=0$. The effect of the mass parameter in this space-time remains open.

 Acknowledgements. Dario Nu\~nez for many helpful discussions. The numerical computations were
carried out in the "Laboratorio de Super-C\'omputo
Astrof\'{\i}sico (LaSumA) del Cinvestav", in the UNAM's cluster Kan-Balam and in the cluster Xiuhcoatl from Cinvestav. This work was partially supported by CONACyT M\'exico
under grants CB-2009-01, no. 132400, CB-2011, no. 166212,  and I0101/131/07 C-
234/07 of the Instituto Avanzado de Cosmologia (IAC) collaboration
(http://www.iac.edu.mx/). GM is supported by a CONACYT scholarships.

\end{document}